%% file: ixp-hist-conext19.tex
\documentclass[sigconf,10pt]{acmart}

\usepackage{fancyhdr}
\fancypagestyle{plain}{%
	\fancyhead{} %
	\rfoot{\today\ at \currenttime}
}
\def\BibTeX{{\rm B\kern-.05em{\sc i\kern-.025em b}\kern-.08emT\kern-.1667em\lower.7ex\hbox{E}\kern-.125emX}}

\setcopyright{none}
\renewcommand\footnotetextcopyrightpermission[1]{}

\usepackage[yyyymmdd,hhmmss]{datetime}
\usepackage{booktabs} 
\usepackage{multirow}
\usepackage[nolist]{acronym}
\usepackage{xspace}
\usepackage[english]{babel} 
\usepackage{hyperref}
\usepackage{url}
\usepackage{xcolor}
\usepackage{subcaption}
\usepackage{placeins}
\usepackage{balance}
\usepackage{graphicx}

\ifx\commentsoff\undefined
	\newcommand\ignacio[1]{\textbf{\textcolor{orange}{IGN: #1}}}
	
\else
	\newcommand\ignacio[1]{}
\fi

\input{./config/acros}

\input{./config/macros}

\makeatletter
\renewcommand\paragraph{
	\@startsection{paragraph}
	{4}
	{\z@}
	{1ex \@plus1ex \@minus.2ex}
	{-1em}
	{\normalfont\normalsize\bfseries\maybe@addperiod}}
\newcommand{\maybe@addperiod}[1]{%
	#1\@addpunct{.}%
}
\makeatother

\begin{document}

	\title{Shaping the Internet: \\ 10 Years of IXP Growth}

\author{Timm B\"ottger}
\affiliation{%
	\institution{Queen Mary University of London}
}
\email{timm.boettger@qmul.ac.uk}

\author{Gianni Antichi}
\affiliation{%
	\institution{Queen Mary University of London}
}
\email{i.castro@qmul.ac.uk}

\author{Eder L. Fernandes}
\affiliation{%
	\institution{Queen Mary University of London}
}
\email{e.leao@qmul.ac.uk}

\author{Roberto di Lallo}
\affiliation{%
	{\institution{\small{Roma Tre Univ, Consortium GARR}}}
}
{\email{dilallo@dia.uniroma3.it}
	
	\author{ Marc Bruyere}
	\affiliation{%
		\institution{University of Tokyo}
	}
	\email{mbruyere@nc.u-tokyo.ac.jp}
	
	\author{Steve Uhlig}
	\affiliation{%
		\institution{Queen Mary University of London}
	}
	\email{steve.uhlig@qmul.ac.uk}
	
	\author{Gareth Tyson}
	\affiliation{%
		\institution{Queen Mary University of London}
	}
	\email{g.tysong@qmul.ac.uk}

	\author{Ignacio Castro}
	\affiliation{%
		\institution{Queen Mary University of London}
	}
	\email{i.castro@qmul.ac.uk}
	\renewcommand{\shortauthors}{F. Lastname et al.}
	
	%
	\begin{abstract}
		\input{./sections/abstract}

	\end{abstract}

	\maketitle
	
	\input{./sections/introduction}

	\input{./sections/data}

	\input{./sections/reachability}
	\input{./sections/hierarchy}
	\input{./sections/related}
	\input{./sections/conclusions}
	\balance
	\bibliographystyle{ACM-Reference-Format}
	\bibliography{references}
	
\end{document}

%% file: config/acros.tex
\begin{acronym}
	\acro{PoW}{Proof of Work}
	\acro{SDN}{Software Defined Network}
	\acro{Tor}{The Onion Router}
	\acro{RIB}{Routing Information Base}
	\acro{P2P}{Peer-to-Peer}
	\acro{BGP}{Border Gateway Protocol}
	\acro{ISP}{Internet Service Provider}
	\acro{PoR}{Proof of Retrievability}
	\acro{IXP}{Internet eXchange Point}
	\acro{CDN}{Content Delivery Networks}
	\acro{QoS}{Quality of Service}
	\acro{SLA}{Service-Level Agreement}
	\acro{ISP}{Internet Service Provider}
	\acro{AS}{Autonomous System}		\acrodefplural{AS}[ASes]{Autonomous Systems}
	\acro{EGP}{Exterior Gateway Protocol}
	\acro{CIPT}{Cooperative IP Transit}
	\acro{T4P}{Transit for Peering}
	\acro{IP}{Internet Protocol}
	\acro{VoIP}{Voice over IP}
	\acro{CDR}{Committed Data Rate}
	\acro{OCR}{Optical Character Recognition}
	\acro{BIX}{Budapest Internet eXchange}
	\acro{FICIX}{FInnish Communication and Internet eXchange}
	\acro{NIX}{Neutral Internet eXchange}
	\acro{SIX}{Slovak Internet eXchange}
	\acro{IIX}{Israeli Internet eXchange}
	\acro{DE-CIX}{German Commercial Internet Exchange}
	\acro{AMS-IX}{Amsterdam Internet Exchange}
	\acro{LINX}{London Internet Exchange}
	\acro{NYIIX}{New York International Internet Exchange}
	
  	\acro{LG}{Looking Glass}
  	\acro{CDF}{Cumulative Distribution Function}
  	\acro{PoP}{Point of Presence}
	\acro{FeRS}{Federated Route Server}
	\acro{PNI}{Private Network Interconnect}		\acrodefplural{PNI}[PNIs]{Private Network Interconnecs}

\end{acronym}

\newcommand{\ixp}{\ac{IXP}\xspace}
\newcommand{\ixps}{\acp{IXP}\xspace}
	\newcommand{\cdns}{\acp{CDN}\xspace}

\newcommand{\as}{\ac{AS}\xspace}	\newcommand{\ases}{\acp{AS}\xspace}



%% file: config/macros.tex
\newcommand{\allnotes}[1]{}
\renewcommand{\allnotes}[1]{\textit{#1}}

\newcommand{\bi}{\begin{itemize}}
\newcommand{\ei}{\end{itemize}}

\newcommand{\eg}{{\it e.g.,}\xspace}
\newcommand{\ie}{{\it i.e.,}\xspace}

\newcommand{\period}{{2008-2016}\xspace}

%% file: sections/abstract.tex
Over the past decade, IXPs have been playing a key role in enabling interdomain connectivity.
Their traffic volumes have grown dramatically and their physical presence 
has spread throughout the world. While the relevance of IXPs is undeniable, their long-term contribution to the shaping 
of the current Internet is not fully understood yet.

In this paper, we look into the impact on Internet routes of the intense IXP growth over the last decade. 
We observe that while in general IXPs only have a small effect in path shortening, very large networks do enjoy a clear IXP-enabled path reduction.
We also observe a diversion of the routes, away from the central Tier-1 ASes supported by IXPs. 
Interestingly,
we also find that whereas IXP membership has grown, large and central ASes have steadily moved away from public IXP peerings, whereas smaller ones have embraced them.
Despite all this changes, we find though that a clear hierarchy remains, with a small group of highly central networks

%% file: sections/introduction.tex
\section{Introduction}
\label{sec:intro}
Originally, the Internet had a hierarchical structure, dominated by a small clique of transit providers (Tier-1s) which guaranteed global connectivity~\cite{labovitz2010internet}. 
In recent years, however, a number of networks have begun to bypass these operators (via direct peering connections), thereby creating a de-hierarchization or ``flattening'' of the Internet structure~\cite{gill2008flattening,dhamdhere2010internet}. 
By reducing interconnection costs~\cite{castro2014using}, switching facilities known as \ixps, have emerged as the default 
location for peering~\cite{ager2012anatomy}. With this surge in peering, \ixps have gone through a startling 
growth: Leveraged by \cdns~\cite{bottger2018open}, the traffic volumes of \ixps have grown dramatically~\cite{chatzis2013there}, and their physical presence has became pervasive, either through new facilities~\cite{chatzis2015quo} or by attracting far away networks via remote peering providers~\cite{castro2014remote,nomikos2018peer}.
%

While previous research has shown the critical impact of a specific \ixp in a concrete moment of time~\cite{ager2012anatomy,will2013vantage}, our community lacks a clear and nuanced understanding of 
their magnitude and long-term impact.
In addition, \ixps are critical infrastructures~\cite{Giotsas2017outages} and are understood to be 
a fundamental bootstrapping step in the development of a region's communication infrastructure\footnote{\url{https://www.internetsociety.org/policybriefs/ixps/}}.
Accordingly, we argue that a comprehensive understanding of their evolution is key for informing future steps. 

In this paper, we investigate the evolution of \ixps and their impact on Internet paths 
over a period covering nearly a decade. We first study how the \ixp ecosystem has evolved by leveraging a 
long set of historical (\period) snapshots from PeeringDB~\cite{lodhi2014using}. 
Then, after identifying \ixps in historical traceroute data from iPlane~\cite{madhyastha2006iplane} and CAIDA Ark~\cite{caida-traceroute}, we examine how \ixps have impacted end-to-end Internet routes. 

We find that the impact of this \ixp surge comes with interesting contrasts. 
For example, on the one hand, while the number of \ixps has tripled over the \period period, 
the address space reachable through \ixps has stagnated. 
For the first time, we disentangle Internet flattening from path shortening, highlighting critical nuances. 
We show that whereas the large \ixp growth has resulted in a significant path-shortening to large global networks (\ie hypergiants~\cite{bottger2018looking}), for the rest of destinations the path-shortening  is rather modest.
We also explore how \ixps have radically flattened the Internet by reducing the dependence on Tier-1 \ases, and how they have reduced the dependency on transit, more generally. 
At the same time, \as composition has changed over time with large and central \ases steadily moving away from public peerings at \ixps, whereas smaller ones have embraced them.
Finally, and despite all these changes, we find that a clear hierarchy still
remains, with a small number of networks playing a central role.

In summary, the main contributions of this paper are: 
\begin{enumerate}
	\item \textit{Evolution of the \ixp ecosystem over the past decade:} we observe a threefold increase in the number of \ixps and members therein, accompanied by a stagnation in the fraction of address space reachable through these facilities. 
	\item \textit{Impact of \ixps on the path lengths:} we find that while \ixps have a modest effect on path lengths over time, they have a decisive impact in the shortening of paths towards very large networks.
	\item \textit{Impact of \ixps on the Internet structure:} we observe how  \ixps have reduced transit dependence of \ases, enabling a clear bypassing of Tier-1 transit providers. We find however that the Internet still has a clear hierarchical structure with a small set of networks remaining highly central. 	
	Simultaneously, we also observe how central and large \ases have steadily moved away from public peerings at \ixps, whereas less central and smaller, ones have embraced them. 
\end{enumerate}

We argue that these results are of critical importance to 
regions (like Africa), which are currently undergoing a strategic expansion of \ixps~\cite{fanou2018exploring} and regard \ixps as a critial element in the process, as well as
\ixp operators wishing to expand or establish new facilities and network operators formulating their own \ixp strategy. As such, we will make our datasets and code open for the community to utilize.

%% file: sections/data.tex
\section{Data and methodology}
\label{sec:data}

We start by describing the datasets used to underpin our subsequent analysis. 

\paragraph{IXPs and membership}
To gain a groundtruth of IXP presence and membership, we exploit about a decade of data from PeeringDB for the years ranging from 2008 to 2016. 
PeeringDB contains a comprehensive snapshot of the \ixp ecosystem~\cite{lodhi2014using,bottger2018looking} 
and operates as a voluntary platform for \ases to 
expose their presence and other relevant information (\eg 
willingness to peer) to facilitate peering interconnections.
We obtained the years 2008 and 2009 using the ``Way 
Back Machine''~\footnote{\url{http://archive.org/web/}} and
years 2010 to 2016 from CAIDA~\cite{caida-peeringdb}. 
A full overview of the methodology is available in~\cite{bottger2018looking}.

\paragraph{Traceroute data}
The above only provides information about the make-up of individual \ixps. 
It does not reveal interconnectivity between neither networks or \ixps. 
Hence, we also gather two large-scale traceroute datasets spanning over a decade: \textit{(1)} The iPlane 
project~\cite{madhyastha2006iplane} includes traceroutes launched from 
PlanetLab nodes to addresses in all the routable prefixes from the year 2006 
to 2016, and \textit{(2)} CAIDA Ark~\cite{caida-traceroute} includes traceroutes from Ark 
monitors to randomly selected destinations for each routed IPv4 /24 prefix 
on the Internet since 2007. iPlane and Ark complement each other: 
while iPlane vantage points are usually located within academic networks, 
Ark traces originate from a wider variety of networks. We use all available data from the iPlane 
project (June 2006 to August 2016) and a similar timeframe (October 2007 to August 2016) 
from Ark.\footnote{There is no Ark data prior to October 2007.}
We aim at using time-aligned snapshots in a monthly fashion. For iPlane we 
use, whenever possible, the snapshots taken on the first of each month. Similarly, 
for Ark, we use all measurement cycles covering the first of each month, or if no snapshot
is available that day, we use the previous or next day if available.
Note that iPlane suffered a large outage with no data available between November 2010 
and July 2011. Other than this, there are only four months for which we could 
not find an iPlane snapshot. As these months are non-consecutive, we expect 
limited impact on the provided results. In the case of Ark, we could find snapshots 
for all months expect one. In total, we use snapshots from 109 different months 
for iPlane and 107 for Ark.

\paragraph{Identification of IXPs}
As we are interested in understanding the role of \ixps, we tag each traceroute with information about any \ixps the path traverses. We identify \ixp crossings with the traIXroute tool~\cite{nomikos2016traixroute}. This labels each path with which \ixp was traversed, and at which hop.

\paragraph{Sanitization of traceroutes.}
The selected snapshots contain more than 6.7 billion individual traceroutes, 2.3 
billion from iPlane and 4.4 billion from CAIDA Ark. To ensure the reliability of our 
results, we followed a three step approach, applying 3 filters on the dataset.

\textit{Step 1: destination IP filter}. To avoid underestimating the length of a path 
in the Internet, we require that each traceroute contains a reply from the destination IP.
This invalidates 3.9 billion Ark traceroutes and 930 million iPlane traceroutes. We also find that 
496 Ark traceroutes and 8.2 million iPlane traceroutes contain more than one reply from the 
destination IP\footnote{These iPlane traceroutes do not terminate, sending unnecessary probes, 
	while we can see an answer from the destination.}, within the same snapshot. We only keep traceroutes 
containing exactly one reply from the destination IP. After applying this filter, we have 503 million 
traces (11.4\%) for Ark and 1.4 billion (60.4\%) for iPlane.

\textit{Step 2: Unresolved hop filter}. We further require that no more than one consecutive hop is 
unresolved. In entering and exiting an \as, a traceroute will typically exhibit (at least) 
two IP hops. Therefore, we consider this unresolved IP hop to prevent cases where this unresolved
hop would actually be an additional \as on the path and would not be accounted for.
This steps discards 141k (3.2\%) and 321k (13.6\%) of Ark iPlane traceroutes respectively.

\textit{Step 3: Single \ixp filter}. Finally, we require that traceroutes do not contain more than 
one \ixp crossing. In conformance with valley-free routing expectations, we assume that 
there should not be more than one peering relationship on a given path. Therefore, for safety, 
in the few cases where we observe more than one \ixp, we discard the traceroute. While valley-free 
routing violations happen~\cite{giotsas2012valley}, this filter eliminates less than 15k 
traces, \ie below 1\% of the traces of each data set.

After the complete sanitization process (summarized in Table~\ref{tab:dataset-sizes}), we 
end up with 358M Ark and 1.1 billion iPlane traces.

\begin{table}
	\centering
	\begin{tabular}{l|rr|r}
		\toprule
		~ & iPlane & Ark & total \\
		\midrule
		All traceroutes & 2.3B & 4.4B & 6.7B \\
		Step 1: Destination IP filter & 1.4B & 503M & 1.9B \\
		Step 2: Unresolved hop filter  & 1.1B & 362M & 1.5B \\
		Step 3: Single \ixp filter & 1.1B & 358M & 1.5B \\
		\bottomrule
	\end{tabular}
	\caption{Sizes of the datasets, before and after the sanitization process.}
	\label{tab:dataset-sizes}	
\end{table}

\paragraph{Relationships between ASes}
To gain further insight into the nature of interconnections between networks, we further annotate each AS-link with its economic relationship. 
We use the classifier developed by CAIDA~\cite{caida-relationships}, which labels each link as either peering or transit.
Whereas a transit provider typically sells access to the global Internet, in a peering relationship, two \ases exchange the traffic of their respective \emph{customer cones}. 
The customer cone of an \as contains all the \ases that can be reached 
through provider-customer relationships, \ie it is the set of customer \ases, and 
their customers in a recursive manner. 
In terms of reachability, the customer cone of a network contains the address space that the \as can reach either directly or through its customer links. 
In particular, we use the \ases' customer cones computed with the methodology 
from~\cite{luckie2013relationships}. We then calculate \ases' reachability 
by mapping announced IP prefixes to the respective \ases and extracting the unique 
set of such IPs.
Note that throughout this paper, we only consider IPv4, as the impact of IPv6 has become tangible only for the
later years considered~\cite{czyz14}.

\paragraph{Reachability of IXPs}
To infer the upper bound of reachability that \ixps provide to \ases,
we calculate the customer cone of each \ixp.
While the concept of customer cone relates to \ases, we 
extend it to \ixps to understand the reachability that can be attained 
through them. We consider the customer cone of an \ixp as the union of 
the customer cones of its members. We also discard Tier-1 
\ases from the customer cone: as Tier-1s avoid peering with non-Tier-1 
\ases, which they regard as their potential customers~\cite{t1syed2012}, 
this is a realistic assumption. 
The reachability attainable through the customer cone of an \ixp
informs of the destinations that a new \as could attain 
by colocating there~\cite{castro2014remote} and peer with all the networks present 
at the facility with the exception of the Tier-1 \ases.

\paragraph{Identification of Tier-1s}
Finally, we label all instances of Tier-1 ASes~\cite{t1syed2012} in our traceroutes (\ie  provider-free \ases that reach the entire Internet without paying other \ases for transit services). 
We consider Tier-1 ASes as the clique of the \as graph as defined in~\cite{luckie2013relationships,caida-relationships}. 
While the announced IP space of these Tier-1 \ases is only about 10\% of all announced IPs, their joint reachability is much larger: 
Tier-1 \ases reach about 99\% of the announced address space, either directly or through their customer cone.

\subsection{Data Coverage}
\label{sec:data-validation}

Although we have gathered, to date, the most comprehensive historical dataset on \ixp evolution, it is important to identify the geographical and network bias that exists in any such dataset. 
We now assesses the coverage and bias of our traceroute datasets, such that readers can interpret our results appropriately.
Specifically, we look at the geographic coverage of the measurement sources as well as their targeted destinations. 





We are confident that our traceroute dataset provides sufficient coverage:
In the following pargraphs, we show that the data collected has a sufficient coverage of the Internet in terms of \textit{(1)} geographical regions, \textit{(2)} prefixes and \textit{(3)} \ases and IPs. We also verify that our sanitation steps do not significantly reduce or bias the coverage provided by the data.



\paragraph{Geographic Probe Coverage}
Before continuing, it is important to quantify the geographical coverage and bias contained without dataset. 
Since PlanetLab does neither encode countries into hostnames directly nor provides historic 
information on decommissioned nodes, we cannot automatically obtain location 
information for all iPlane nodes. Where available, we instead use the Top Level Domain (TLD) of the 
PlanetLab node name to infer the country where the node is/was located (\eg
\textit{planet-lab1.itba.edu.ar} or \textit{planetlab1.itwm.fhg.de}). This is 
feasible for 657 (54\%) PlanetLab nodes. Out of the remaining 561 PlanetLab nodes, 
a majority of 384 domain names map to the .edu TLD, which in most cases
maps to US universities. In cases where we could not rely on the TLD, we resolve 
names to countries manually. While the resolution in most cases was 
straightforward (\eg \textit{planetlab-1.cs.princeton.edu}), it sometimes
required more manual inspection (\eg \textit{plab1.create-net.org} appears to be 
an Italian node). We sanitized data to the best of our knowledge, but cannot rule out small issues. Nevertheless, we have high confidence in at least the mapping of country-specific TLDs and the .edu TLD, which corresponds to more 
than 85\% of all node names. As for Caida Ark, the mapping of nodes to 
countries could be derived automatically, since all Ark nodes embed the 
TLD of the countries in the node name.

Table~\ref{tab:probes-continents} shows the distribution of measurement nodes with
respect to continents: Europe, Asia and Africa dominate. However, to 
put values into the right perspective, in the total column we also denote how many 
countries of a continent are covered by iPlane and Caida Ark and provide the respective 
percentages in the table  (column ``\% Cnt''). We see that while Europe is well covered, in terms 
of relative coverage, South America is performing better than Africa and Asia. To get 
a better understanding of which countries within a continent are covered, we also assess which 
fraction of a continent's population the covered countries represent (column ``\% Pop.'').
At least half of the population in each continent has a measurement probe 
deployed in their country. Specifically, North America and Asia achieve more than 80\% 
coverage, followed by South America and Oceania with more than 70\% each. Europe achieves 
more than 60\% coverage while only Africa gets less than 60\% coverage. Coverage 
appears to be in line with the development status of continents, with a bias towards 
better coverage for more developed regions. 


\begin{table}
	\centering
	\begin{tabular}{l|rr|r|r|r}
		\toprule
		Continent & iPlane & Ark & Total & \% Cnt. &  \% Pop. \\
		\midrule
		Africa & 2 & 14 & 16 of 58 & 27.6\% & 53.1\% \\
		Asia & 13 & 12 & 17 of 58 & 29.3\% & 82.9\% \\
		Europe & 22 & 23 & 27 of 57 & 47.4\% & 66.1\% \\ 
		North America & 5 & 4 & 5 of 43 & 11.6\% & 87.7\% \\
		Oceania & 2 & 2 & 2 of 27 & 7.4\% & 73.8\% \\
		South America & 6 & 3 & 6 of 14 & 42.9\% & 76.1\% \\
		\bottomrule
	\end{tabular}
	\caption{Countries per continent in which a probe is hosted for both platforms.}
	\label{tab:probes-continents}
\end{table}



\begin{figure*}
	\begin{subfigure}{\columnwidth}
		\includegraphics[width=\columnwidth]{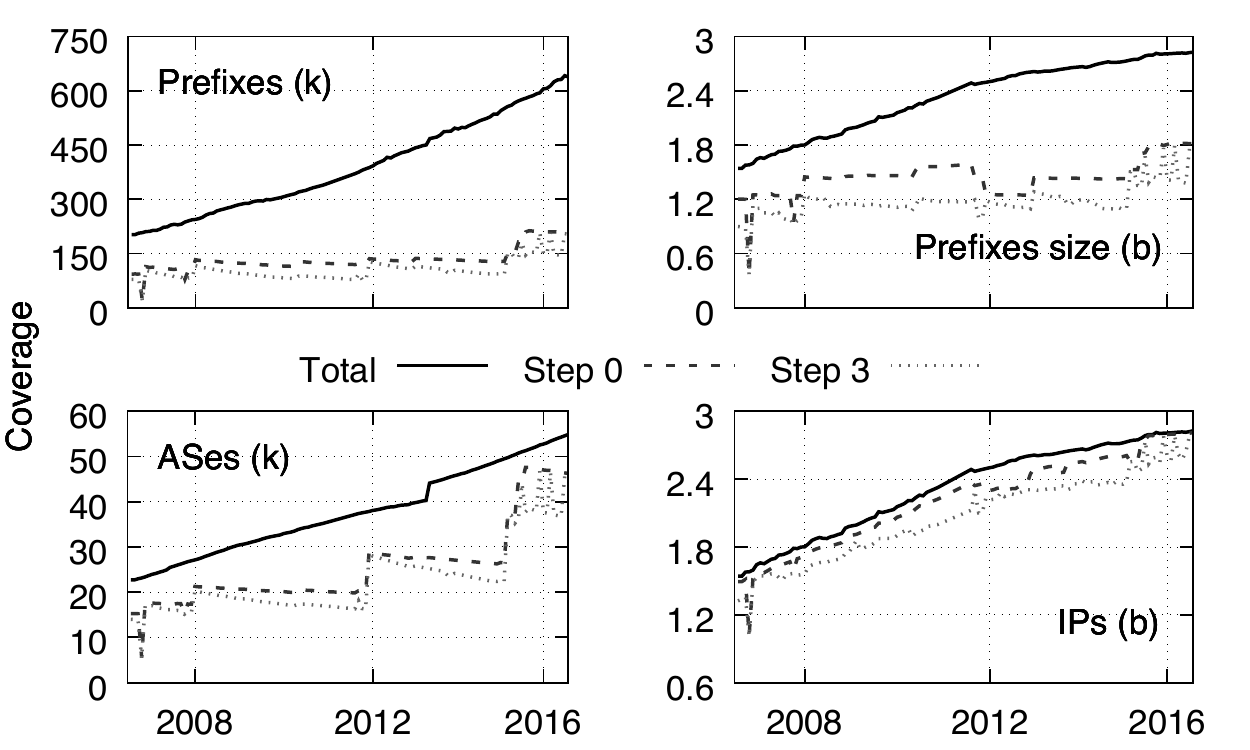}
		\caption{iPlane}
	\end{subfigure}
	\begin{subfigure}{\columnwidth}
		\includegraphics[width=\columnwidth]{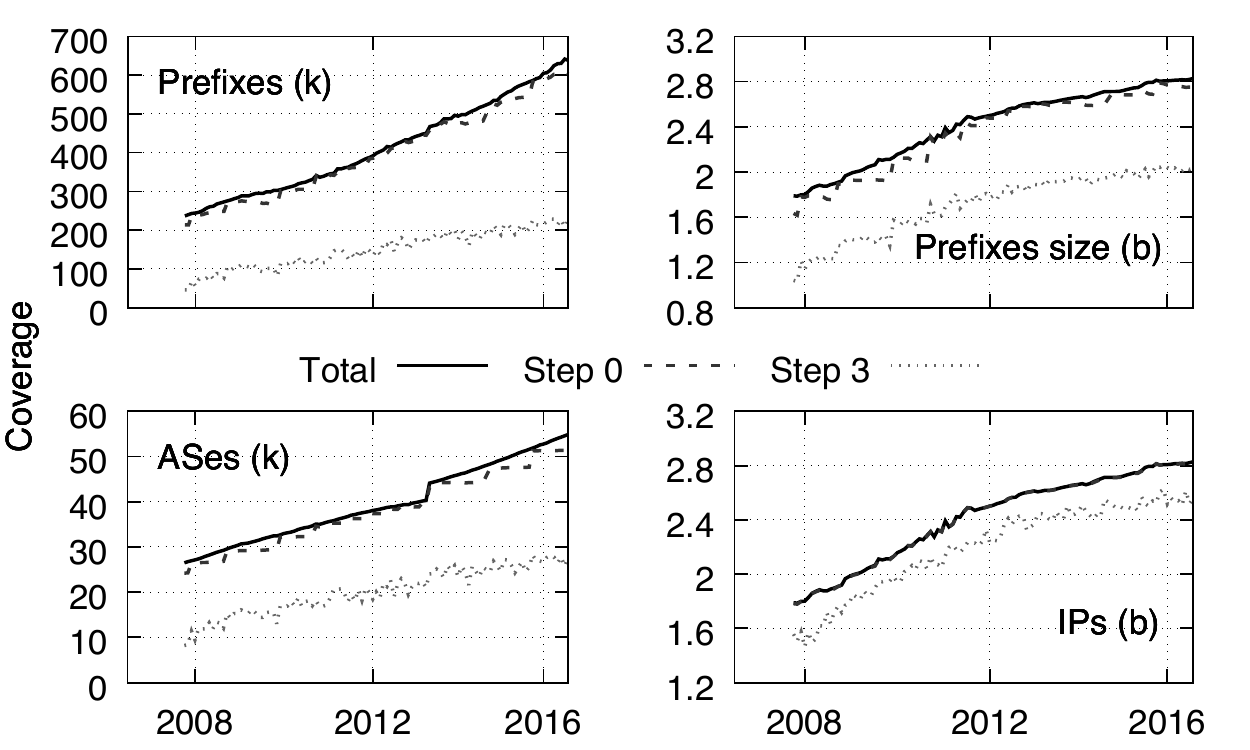}
		\caption{Caida Ark}
	\end{subfigure}
	\caption{Number of prefixes and ASes  (in thousands) and prefix size and the number of IPs (in billions), in the global routing table (\textit{``Total''}), in each dataset (\textit{``Step 0''}) before sanitization, and after sanitization (\textit{``Step 3''}).}
	\label{fig:bias}
\end{figure*}

\paragraph{Prefix Coverage}
To determine which part of the Internet is covered by our traceroute dataset, we first look at 
the prefixes targeted by the traceroutes. Taking as reference all the prefixes in the global routing 
table, as extracted from CAIDA's routeviews data~\cite{caida-routeviews}, we consider only those prefixes not 
longer than /24 -- in accordance with the best BGP practice, that discourages the usage of longer prefixes. 
We count the number of matching prefixes, \ie all the prefixes matching the destination IP of a specific traceroute, 
not just the most specific one.

Figure~\ref{fig:bias} (top four plots) shows for \textit{(1)} iPlane and \textit{(2)} Ark, the number of prefixes (in thousands) and the size of the IP space those prefixes represent (in billions of IPs).
The prefixes of the global routing table are marked as \textit{``Total''}.
The prefixes covered by the original traceroute datasets (\textit{``Step 0''}) are also shown as well as the prefixes that remain after applying the 3 sanitization filters mentioned above (\textit{``Step 3''}).
While iPlane on average only covers 36\% of the announced prefixes, Ark on average achieves a coverage of 96\% of all announced prefixes.
For iPlane the coverage drops to 29\% after sanitization, whereas for Ark due to the bigger number of traceroutes removed, coverage drops to 36\%.
Overall, the two platforms together cover roughly 50\% of the total prefixes, indicating some overlap in the target selection (not shown in the plots).



The prefixes targeted by iPlane and Caida Ark account for 63\% and 97\% of the announced IP space respectively. 
We reach this percentage by considering the prefix coverage and comparing with the total number of IP addresses announced in the global routing table.
After our sanitization, coverage drops to 52\% and 63\% respectively.
However, when combined, the two platforms still probe prefixes accounting for roughly 80\% of the announced IP space.


\paragraph{AS and IP Coverage.}
We now examine our coverage in terms of \ases and IPs.
We consider an AS to be covered if at least one of its IPs is targeted by a traceroute. We map IPs to \ases using~\cite{caida-routeviews}.
Similarly to the previous case, Figure~\ref{fig:bias} (bottom four plots) shows for (a) iPlane and (b) Ark, all the \ases (in thousands) and the IPs (in billions) present in the global routing table (\textit{``Total''}), the \ases and IPs seen by the corresponding traceroute dataset (\textit{``Step 0''}), as well as the \ases and IPs remaining after the sanitization process (\textit{``Step 3''}).
The two platforms independently cover 50\% to 70\% of the total ASes, while their union covers more than 80\%. 
As for the previous case, we also investigated the amount of the IP address space that the covered ASes represent. Specifically, we found that the ASes targeted by both platforms announce more than 90\% of the global IPv4 space.


In the following, unless stated otherwise, we will always only show results obtained from the CAIDA Ark measurements in all Figures and calculations.
In all the analysis steps we carry out, the two platforms consistently yielded results so similar, that for the sake of clarity in the plots, we decided to only show one of the platforms.
We believe that this similarity in the results obtained through two different, independent measurement platforms strongly reinforces the validity of the results.

\subsection{Dataset Considerations}\label{sec:discussion}
Finally, we wish to briefly highlight some key considerations that readers should keep in-mind when interpreting results. First, it is well known that observing the Internet completely and accurately at the same time is not possible~\cite{willinger2013internet}.
We mitigate this by relying on two large historical datasets of traceroutes with different sets of geographically distributed vantage points~\cite{huffaker2012internet}. Whereas iPlane relies on vantage points located within academic networks, Ark probes are distributed over a wider variety of network types, and our analysis confirms that most of the Internet is observable through the considered traceroutes: our dataset includes more than 80\% of the existing \ases, representing  90\% of the global IPv4 space. That said, we wish to emphasize that we cannot draw conclusions beyond the networks and geographical regions we have previously described. Of course, we also note that topology information only offers one perspective into the emerging role of \ixps, as it is agnostic to traffic volumes.

%% file: sections/reachability.tex
\section{Motivation: IXP growth}
\label{sec:reachability}

To motivate the need to understand the impact of \ixps, we begin by exploring their growth and nature using our PeeringDB dataset.



\paragraph{Membership and facilities.}

\begin{figure}
	\centering
	\includegraphics[width=\columnwidth]{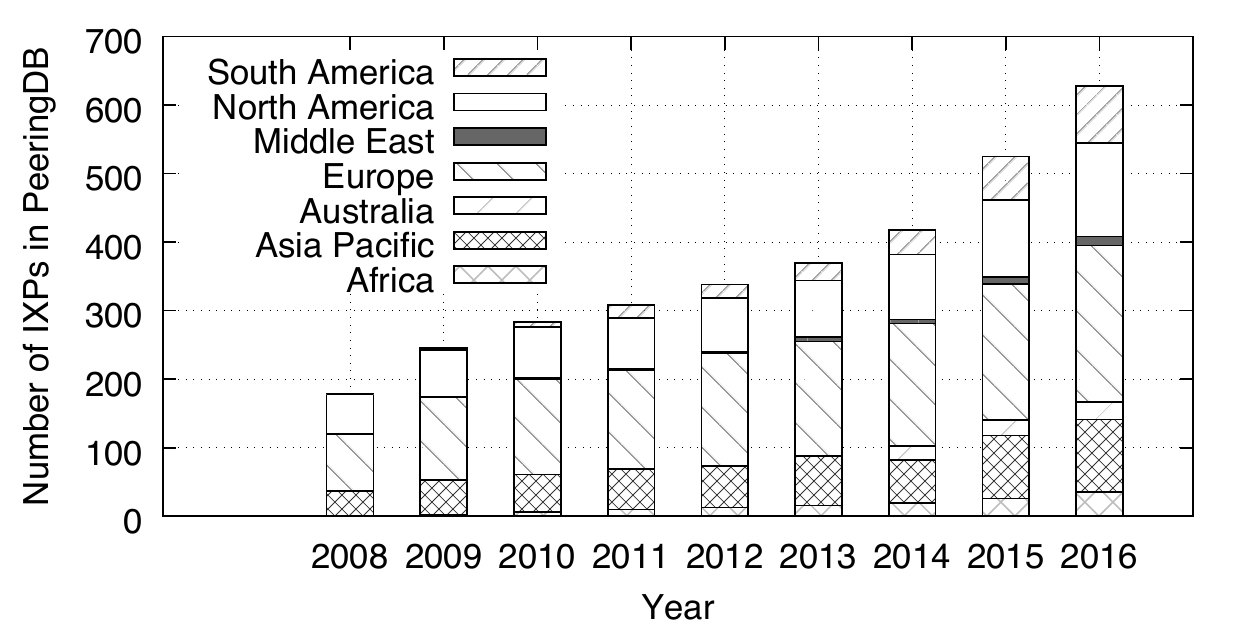}
	\caption{Evolution of the number of IXPs.}
	\label{fig:ixps-year-region}
\vspace{.5cm}
	\includegraphics[width=\columnwidth]{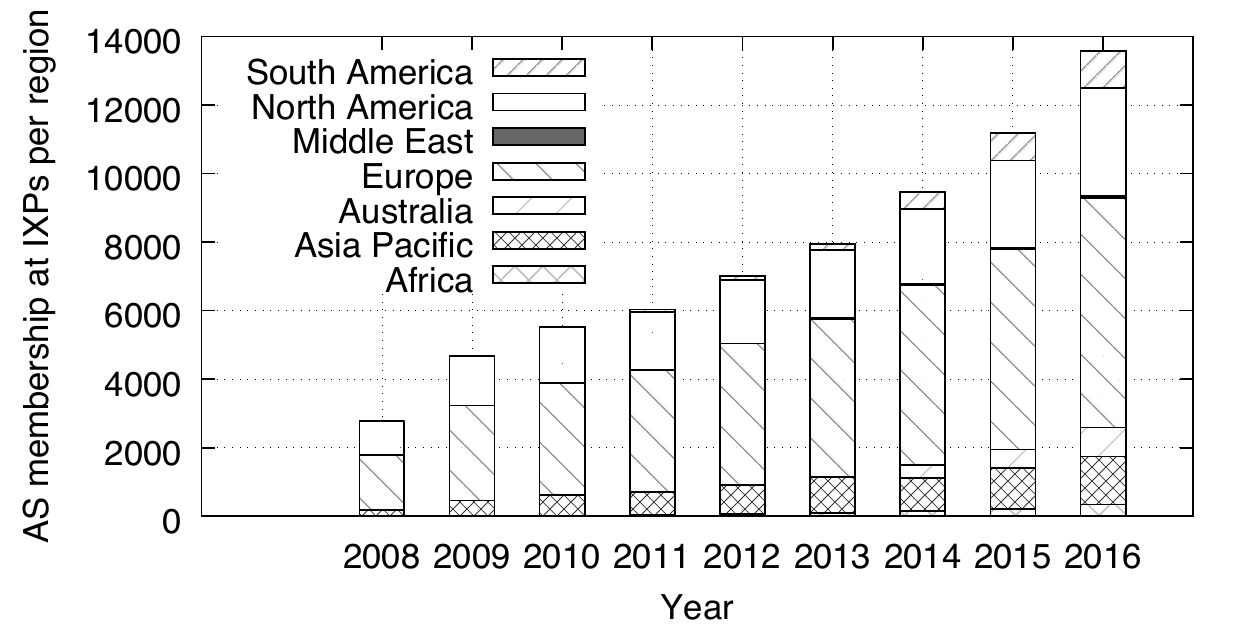}
	\caption{Evolution of the number of IXP members.}
	\label{fig:ases-year-region}
\end{figure}

We first look at the evolution of the number of \ixps across regions within PeeringDB, shown in Figure~\ref{fig:ixps-year-region}.
A decade ago, almost all of the roughly 200 existing \ixps were located in Europe, North America and Asia. 
Nowadays, the number of \ixps has tripled and new \ixps have emerged all over the world.
They  have particularly gained presence in South America and Africa.
Despite the significant growth across all regions, Africa
remains under-represented and Europe remains the most popular region for \ixps.

To gain insight into the evolution of the size of these increasingly numerous \ixps, we now look at the number of \ixp members (peers) per region in Figure~\ref{fig:ases-year-region}.
The proliferation of \ixps has been mirrored by a similar increase in the number of \ases present at \ixps throughout the world.
While in 2008, IXPs in Europe and North America had 1,597 and 1,005 members respectively, they had 6,697 and 3,144 members in 2016.
IXPs have been particularly popular in Europe, where the largest \ixps (measured by number of members) are located.
From 2006 until 2014, the biggest IXP in Europe has always had more than twice as many members as the biggest IXP outside Europe.
Although over the last two years, this imbalance has reduced, the biggest IXP in Europe still has more than 1.8 and 1.4 times as many members as the two biggest IXPs outside of Europe, respectively. This expansion has resulted in a general trend towards increasing average sizes of \ixps in all regions. Nevertheless, in the most recent snapshot, Europe, North America and Australia (with more than 28 members per IXP on average) are still well ahead of the rest of the world (where the average size of \ixps is not greater than 21 members).

\paragraph{Reducing transit dependency.}
Since one of the critical roles of \ixps is to reduce transit dependency, we now investigate to what extent  \ixps have created a potential for bypassing transit providers an in particular T1s.
We measure this via the \emph{reachability} that \ixps can offer without the assistance of Tier-1s, as formalized in  Section~\ref{sec:data}.
We express \ixp reachability as a percentage of the reachability attainable through Tier-1 transit providers. 
Note that as discussed in Section~\ref{sec:data}, this is about 99\% of the announced address space.

\begin{figure}
	\centering
	\includegraphics[width=\columnwidth]{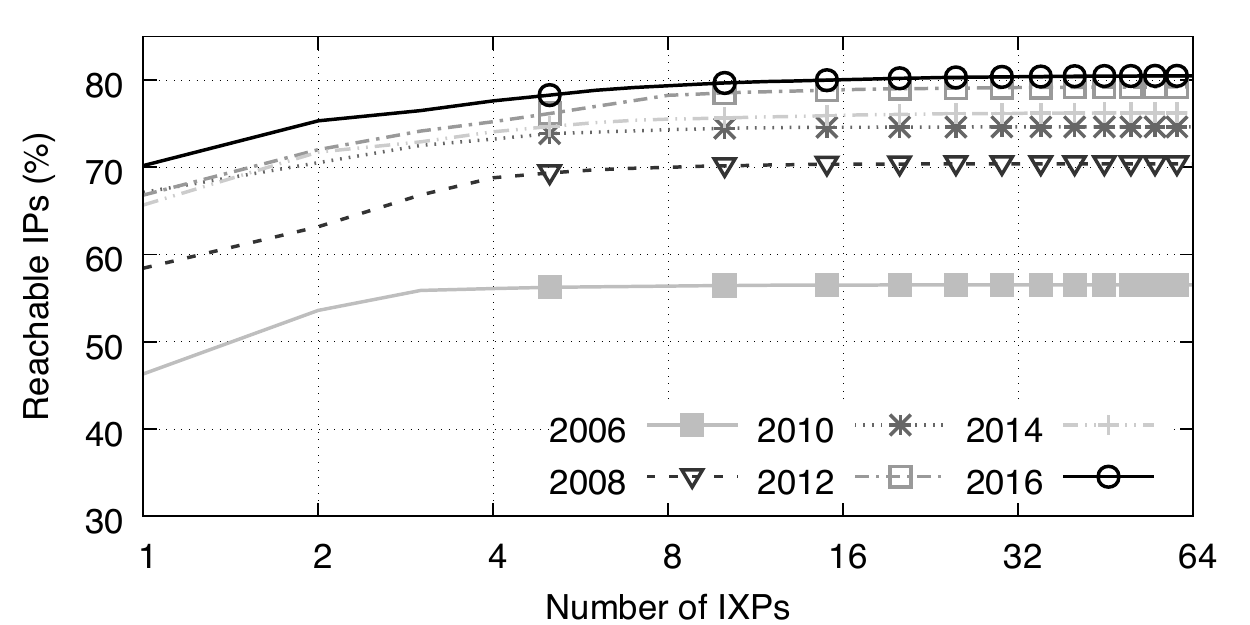}
	\caption{Cumulative reach through IXPs without Tier-1s. IXPs are added iteratively for each year.}
	\label{fig:ixp-reachability-not1}
\end{figure}

Figure~\ref{fig:ixp-reachability-not1} shows the evolving share of Internet destinations that can be reached through \ixps (thereby replacing the need of transit providers). We compute this by iteratively adding the \ixp that provides access to the largest number of IPs which were not reachable so far. While we follow this procedure with all the \ixps in the dataset, we only show the first 64 \ixps (since the marginal contribution is diminishing rapidly). 
The analysis reveals that \ixp growth comes with some interesting nuances.
On average, only the first 25\% of the \ixps (ranging from 35 in 2006 to 114 in 2016) provide additional reachability.
While we observe a dramatic growth in \ixp reachability from just above 55\% in 2006 to about 70\% in just 2 years, this growth comes soon to a halt.
Also, the maximum reachability attainable through \ixps does not grow linearly with time. For instance, 2012's reachability is greater than later years. This is explained by the large growth of Tier-1's reachability after 2012, in conjunction with a stagnation of \ixp reachability --in absolute numbers. This shows that there is still a key role for Tier-1s, albeit one whose prominence is diminishing. 

\begin{figure}
	\includegraphics[width=\columnwidth]{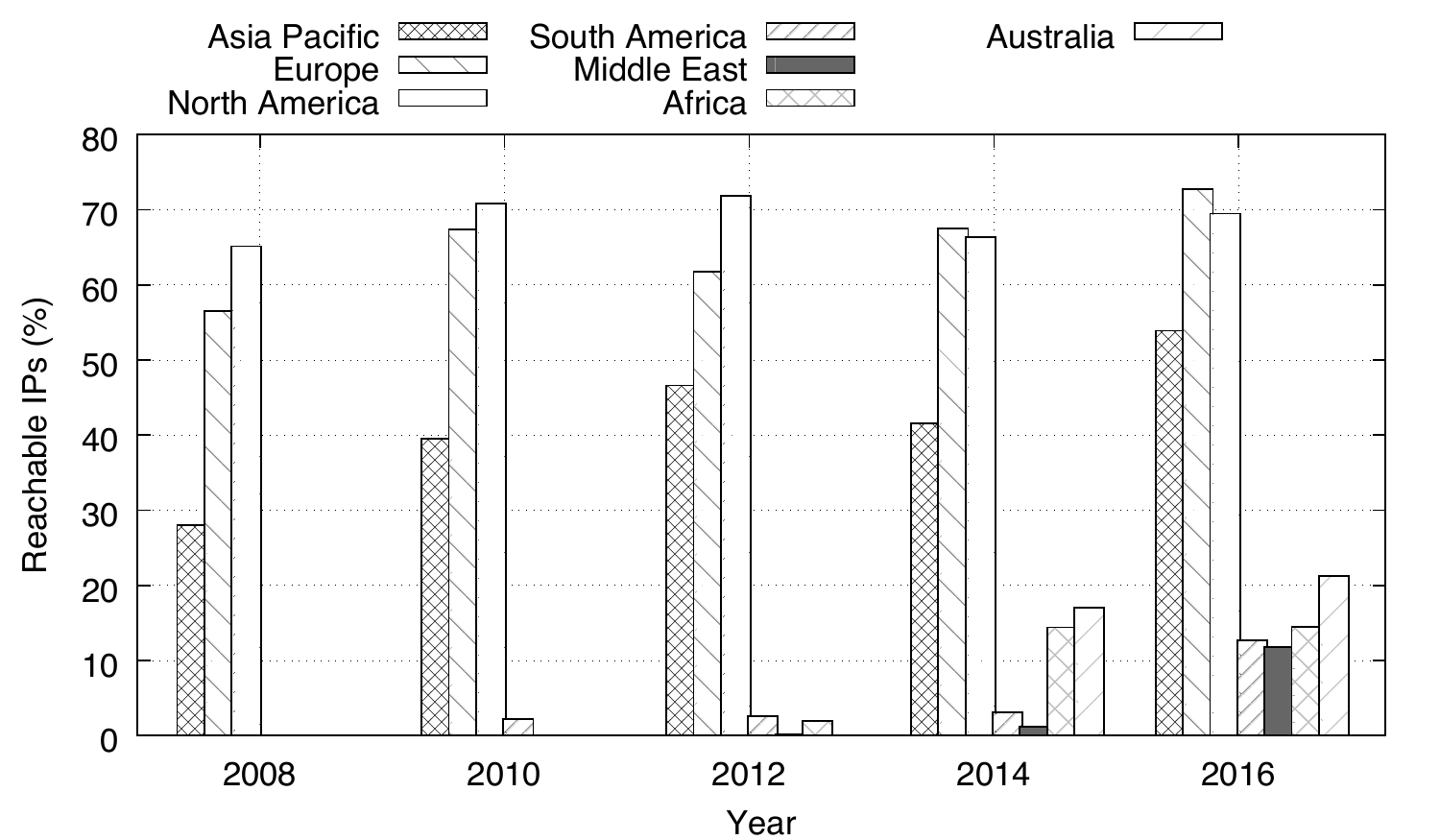}
	\caption{Evolution of the total reach of the IXPs in each region.}
	\label{fig:ixp-reach-year-region}
\end{figure}
\begin{figure*}
	\centering
	\includegraphics[width=1.02\textwidth]{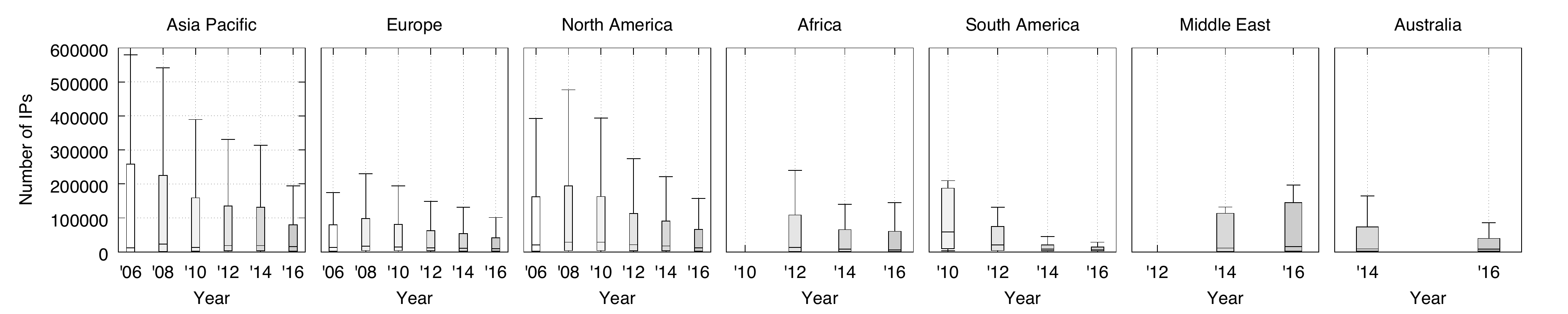}
	\caption{Boxplots depicting the sizes (in terms of announced IPs) of the ASes at any IXP of the given region.}
	\label{fig:boxplot-all-regions}
\end{figure*}

\paragraph{Regional dynamics.}
The above analysis shows that, from a geographical perspective, all regions have grown in terms of \ixp reachability. 
We next dig deeper into this observation to highlight further key nuances.  Figure~\ref{fig:ixp-reach-year-region} presents the total reachability attainable by the \ixps of each region per year, as a percentage of the address space reachable through Tier-1s (\ie similar to Figure~\ref{fig:ixp-reachability-not1}).

Both Europe and North America attain most of the total \ixp-reachability for each year through their local \ixps, with their reachability stabilising at around 70\%.
Interestingly, even though Europe has always had more members and facilities than North America, Europe had lower \ixp reachability. 
This imbalance was only reversed in 2014 due to the larger growth of the European \ixps.
\ixp reachability is therefore not proportional to the size of the \ixp ecosystem.
In contrast, the other regions exhibit a \emph{clear} pattern of growth, with Asia-Pacific well ahead, reaching about 50\% of the address space, while the reachability of the other regions only takes off towards the end of the considered period.
The Asia-Pacific case is particularly striking: despite the small number of \ixps and members, \ixp reachability in Asia is disproportionally high.
This points to a different composition in the \ixp membership across regions.

To gain further insight into the regional dynamics, we look at the size of the \ixp members within each region in terms of the announced address space. Figure~\ref{fig:boxplot-all-regions} presents boxplots with the yearly distribution of \as-sizes for the \ixps in each region. 
We find that while regions are strikingly different, there is a general trend towards smaller \ases.

Throughout almost every region we observe a reduction in the dominance of large networks and  increasingly richer ecosystems formed by more diverse and smaller \ases. 
This results from the shrinking size of the  \ases in the \ixps of almost every region as well as the growing \ixp reachability experienced in each region (see Figure~\ref{fig:ixp-reach-year-region}).
Despite this global trend, the regions are strikingly different:
whereas the \ixp members in  Europe are relatively small, those in North America and Asia-Pacific are much larger. 
Taking into account that the European and North American IXPs had a similar reachability (larger than Asia-Pacific) this points again to a large number of smaller but more diverse \ases present at the European \ixps. 

\begin{figure}
	\centering
	\includegraphics[width=\columnwidth]{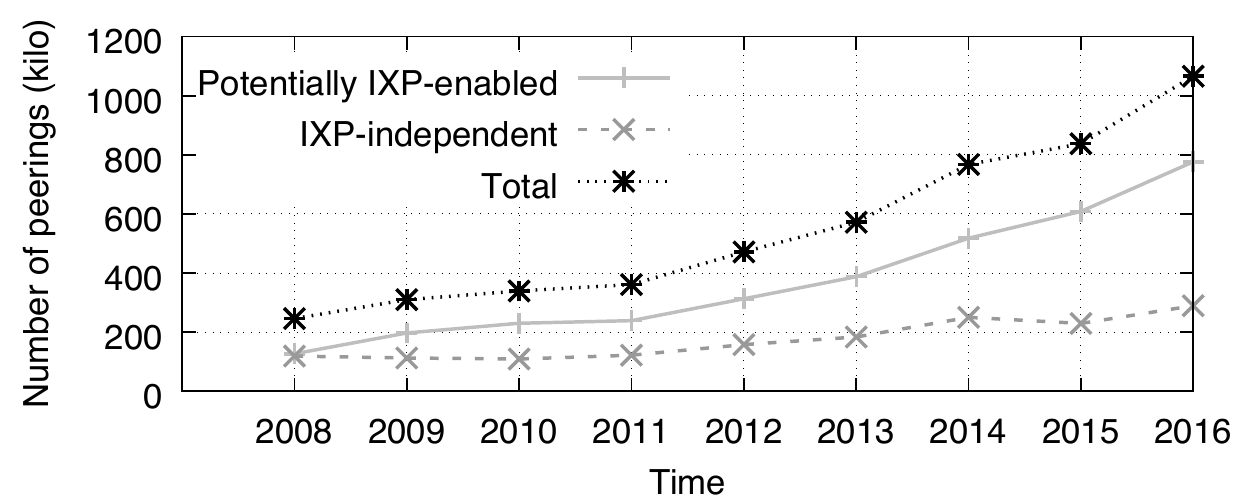}
	\caption{Evolution of the number of potential peerings at and outside IXPs.}
	\label{fig:in-vs-out-ixp-peerings}
\end{figure}

\paragraph{Peerings at \ixps.}
Finally, to understand the potential impact of \ixps on the Internet routes, we now look at the evolution of the number of potential peerings that these facilities can enable.
We do so by inferring an upper bound of \textit{potential} \ixp-enabled peering connections.
We consider a peering as potentially \ixp-enabled when the following two conditions are met:
\textit{1)} \ases are peering, and
\textit{2)} \ases are also colocated simultaneously at the same \ixp (according to PeeringDb).
Figure~\ref{fig:in-vs-out-ixp-peerings} confirms that peering has grown dramatically. Peering has roughly quintupled over our measurement period, primarily due to the the rise in potentially \ixp-enabled peerings. Unfortunately, however, this \emph{potential} upper-bound does not confirm to what extent networks are actually exploiting this opportunity. Hence, in the next section we turn to our traceroute datasets to understand how \ixps have enabled direct connections among networks.


%% file: sections/hierarchy.tex
\section{Impact of IXPs on the Internet}\label{sec:flattening}
%
Whereas we have so far observed the growth of \ixps, we are yet to explore their role in interconnecting global networks, as well as the impact this may have on Internet paths. 
Exploiting our traceroute datasets, we next examine whether \ixps have led to a reduction in path lengths and how they have affected the Internet structure more generally. We do this from two main perspectives: \emph{(1)} Path Length: how many AS and IP-level hops exist on a path; and \emph{(2)} Path Composition: what types of ASes sit on that path. 


\begin{figure}
	\centering
	\includegraphics[width=\columnwidth]{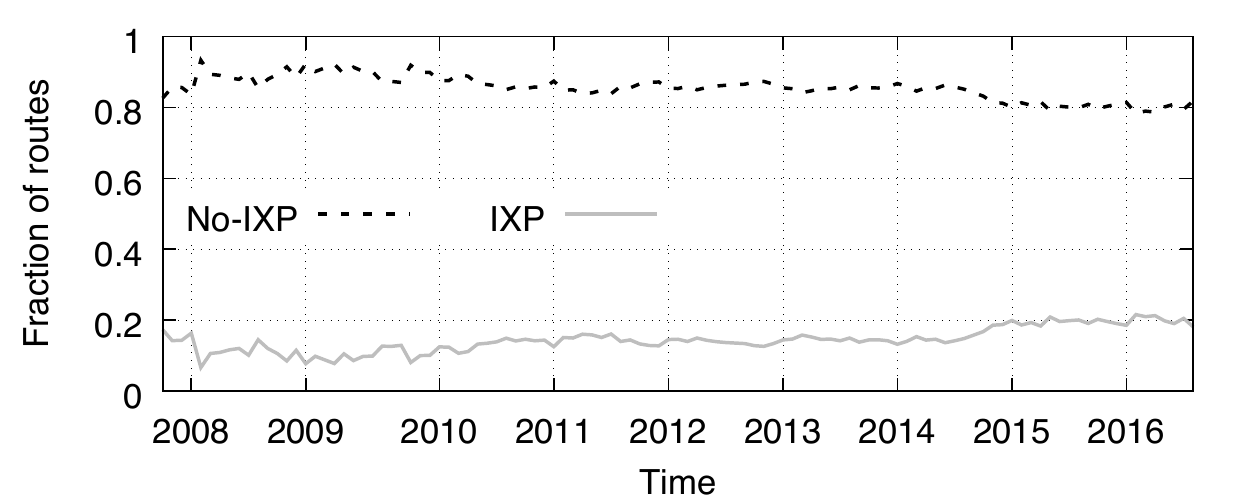}
	\caption{Evolution of the relative share of traceroutes depending on whether they traverse an IXP.}
	\label{fig:traceroute-ixp-non-ixp-filtered-relative-share}
\end{figure}

\subsection{Paths through IXPs}
We first show how prevalent \ixps are on Internet paths.
Figure~\ref{fig:traceroute-ixp-non-ixp-filtered-relative-share} shows the evolution of the relative share of traceroutes that traverse \ixps.
We observe a linear, albeit not dramatic, trend towards a larger fraction of traceroutes going via IXPs.
While in 2008 only about 6\% of the traces in a snapshot traversed an IXP, by 2016 the share had more than tripled to about 20\%.\footnote{The careful reader will have noticed that the behaviour in 2007 might suggests that in that period the share of traceroutes going through IXPs has actually decreased. When inspecting this period before 2008, it appears that it is actually the number of non-IXP traceroutes that has dropped in absolute value, likely due to a measurement artifact, therefore showing up as an increase of IXP traceroutes.}
While this might seem a relatively moderate change, a more than threefold increase in the number of traces traversing an IXP is a strong indicator that IXPs are becoming a critical interconnection facility in today's Internet.

\subsection{Path shortening?}

\begin{figure*}
	\begin{subfigure}{\columnwidth}
		\includegraphics[width=\columnwidth]{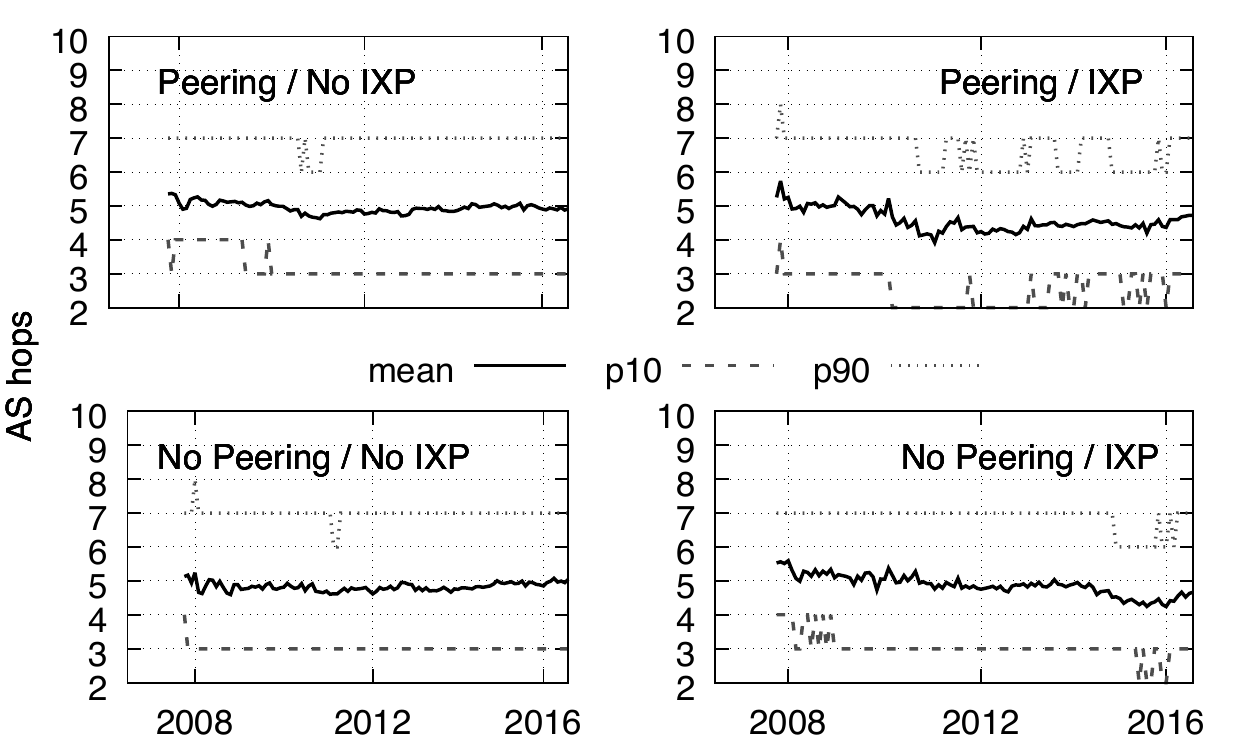}
		\caption{AS-level path-length.}\label{fig:avg-as-hops}
	\end{subfigure}
	\begin{subfigure}{\columnwidth}
		\includegraphics[width=\columnwidth]{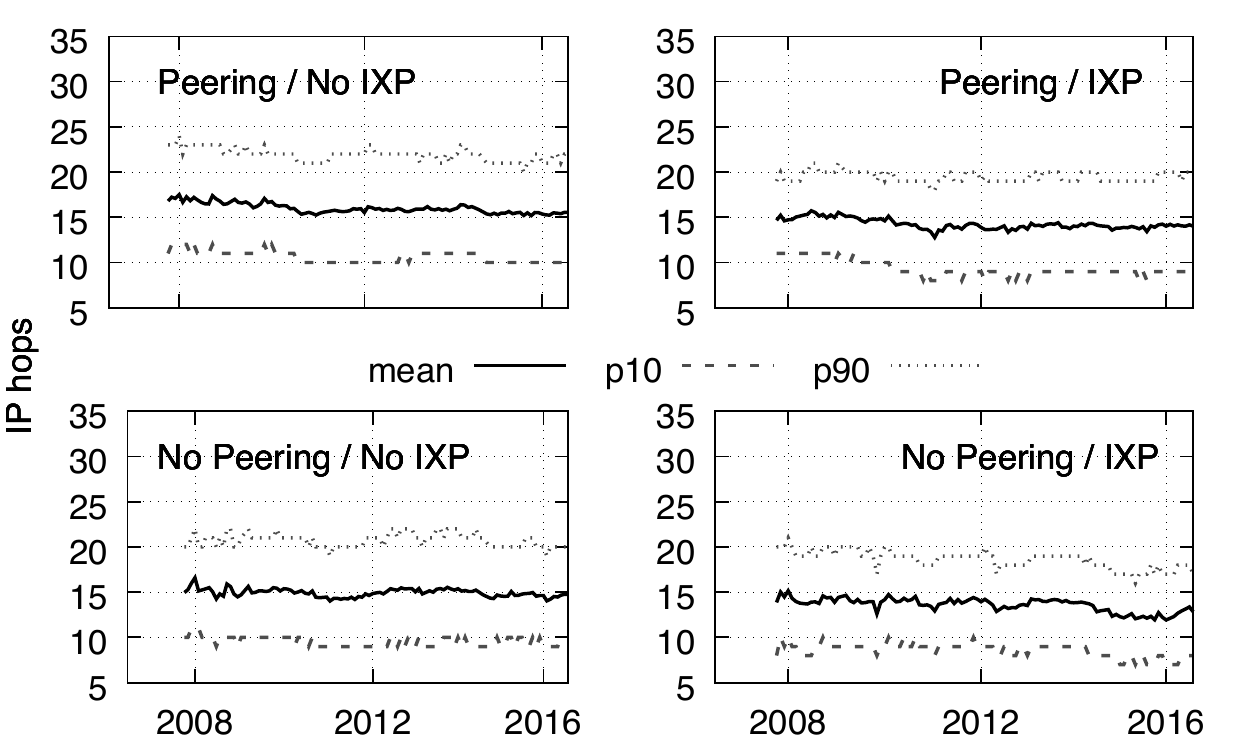}
		\caption{IP-level path-length.}	\label{fig:avg-ip-hops}
	\end{subfigure}
	\caption{Evolution of path-lengths on the AS- (left) and IP-level (right). Data is subgrouped by whether a peering link and/or an IXP is crossed. Metrics shown are mean, 10th and 90th percentile.}
\end{figure*}


One might anticipate that the growing fraction of paths that traverse \ixps would have an impact on average path lengths. 
Note that for many services (\eg interactive media) path length may be a critical consideration for performance~\cite{formoso2018deep}.
While there is a general understanding that \ixps help shorten paths~\cite{di2015really,dhamdhere2010internet,fanou2015diversity}, we are not aware of a thorough longitudinal analysis.


\paragraph{Overall impact}
To assess the evolution of path lengths, we aggregate all the traceroutes by date, measurement system, whether they traverse an IXP and whether they contain a peering link. We then compute the mean, 10th and 90th percentiles of the IP and AS-level path length for each group.
We include the percentiles to check the distributional properties of the paths lengths.
The resulting evolution for both IP and AS hops is depicted in Figures~\ref{fig:avg-as-hops} and~\ref{fig:avg-ip-hops}, respectively. 
As the results for the CAIDA Ark and iPlane were almost identical, we only show the former. The reason for this 4-way breakdown of the path lengths is that one might expect  \emph{(1)} paths going through IXPs to be shorter than those not going through IXPs, and \emph{(2)} paths that neither cross an IXP nor contain any peering to be longer since they are more likely to follow the traditional Internet hierarchy.

In contrast to expectations, the figures show that the number of \as-hops for those traces containing peering links, but no \ixp crossings, has remained mostly \emph{stable} across our measurement period (around 5 hops).
That said, our results confirm that, indeed, \ixps \emph{have} resulted in shorter path lengths, both at the IP and the \as level.
At the IP level, we observe a minor reduction in path-length by about one hop over the period of our traceroute measurements. 
We also observe that traces going through an \ixp are about two IP hops shorter than those crossing no \ixps. 
At the AS level, the increasing number of \ixp peerings has resulted in shorter \as-level paths: they have reduced their length in about 0.5 hops on average. Thus, overall, \ixps have had a clear impact in reducing path-lengths, though the effect is moderate. 
That said, paths that do not contain a peering and do not cross an \ixp are no longer than those with peering but no \ixp.


 


\paragraph{Impact on hypergiants.}
\begin{figure}
	\centering
	\includegraphics[width=\columnwidth]{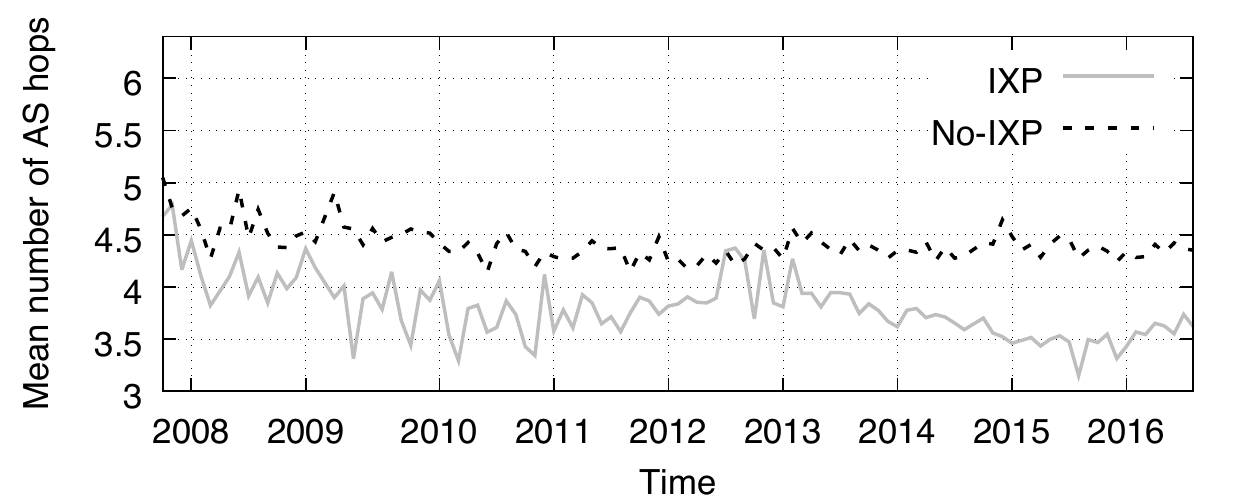}
	\caption{Evolution of the AS-level average path-length to hypergiants. Evolution is shown for traces traversing an IXP and traces traversing no IXP separately.}
	\label{fig:path_length_hypergiants}
\end{figure}
The above reveals a more nuanced understanding of how paths have shortened over time. In contrast, one might expect a more clear-cut case for the shortening of paths towards hypergiants~\cite{bottger2018looking} (\eg Google, Netflix), as these operators actively publicize their desire to peer directly with third party networks. This expectation seems reasonable~\cite{chiu2015we}, as it is those hypergiants that seem to benefit most from the peering opportunities offered by IXPs. 

To explore this, we consider hypergiants as the set of networks identified in~\cite{bottger2018looking}, and calculate their path lengths and average across them for each snapshot.\footnote{The hypergiants are Apple, Amazon, Facebook, Google, Akamai, Yahoo, Netflix, Hurricane Electric, OVH, Limelight, Microsoft, Twitter, Twitch, Cloudflare, Verizon Digital Media Services.}
Figure~\ref{fig:path_length_hypergiants} presents the average number of hops to these hypergiant \ases across our measurement period. Indeed, we observe that \ixps have had a critical role in substantially reducing the paths to hypergiants. 
We find that this impact is twofold:
\textit{(1)} routes traversing an \ixp are typically shorter than those that don't,
\textit{(2)} routes traversing \ixps have enjoyed a path shortening from above 4.5 to about 3.5 \as-hops, whereas the routes with no \ixp have remained roughly stable.
This is reasonable as hypergiants like Netflix and Google are known for their intensive use of IXPs and willingness to peer~\cite{bottger2018open}. 
As these networks are typically responsible for large traffic volumes~\cite{labovitz2010internet, bottger2018open,trevisan2018five}, we conjecture that \ixps are responsible for a substantial path-length reduction for a large fraction of the Internet traffic.

\begin{figure}
	\centering
	\begin{subfigure}{\columnwidth}
		\includegraphics[width=\columnwidth]{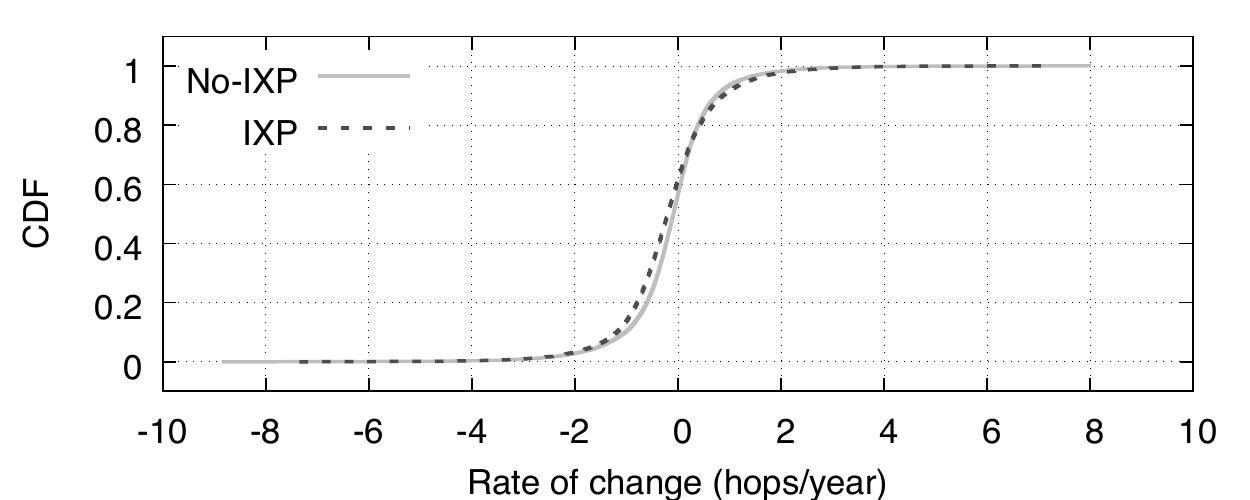}
		\caption{Rate of change of average number of IP-hops per year.}
		\label{fig:linreg-slope-ip-hops-mean-cdf}
	\end{subfigure}
	\begin{subfigure}{\columnwidth}
		\includegraphics[width=\columnwidth]{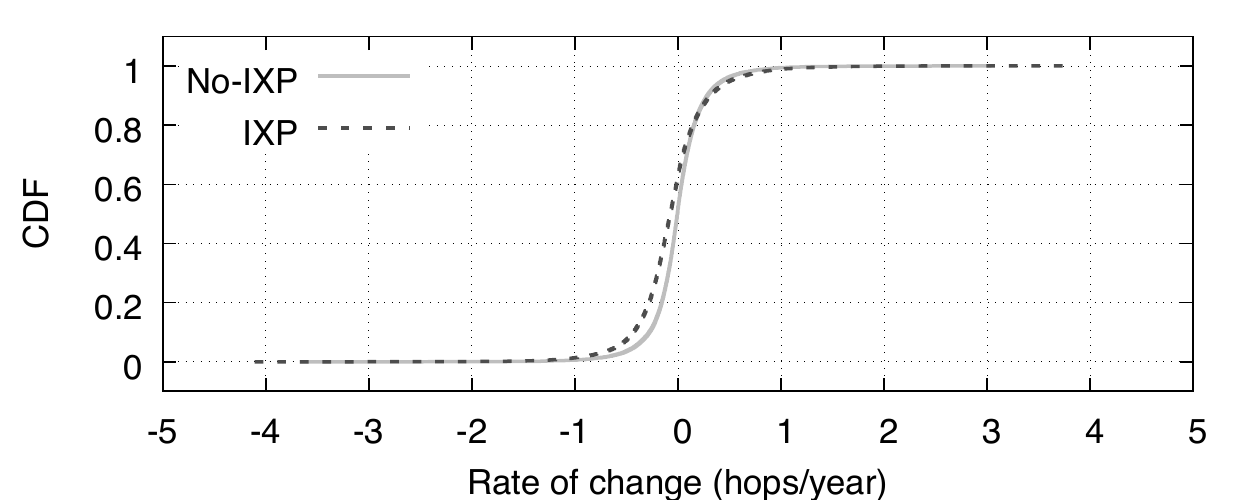}
		\caption{Rate of change of  average number of AS-hops per year.}
		\label{fig:linreg-slope-as-hops-mean-cdf}
	\end{subfigure}
	\caption{CDFs depicting rates of change of the path-length per year per AS, obtained through least-squares linear regression over the average per AS path-length per snapshot.}
\end{figure}


\paragraph{Per-AS path lengths.}
The above has revealed \ixps have lead to paths shortening, particularly for routes to hypergiants. However, it does not shed light onto whether specific AS-to-AS paths are shortening more generally. Hence, in contrast to the previous sections, we next analyse the changes witnessed for a given AS across all snapshots (rather than averaging across all ASes within a snapshot).
We take an AS-centric perspective by computing the average IP and AS-level lengths of all paths for all ASes targeted by the traceroutes and over all measurement snapshots. For each AS, we compute a least-squares linear regression over all mean path lengths over the duration of all snapshots. In this way, we can interpret the slope of the regression line as a linearly flattened rate of change of path length for each AS. A negative rate will then point to a shortening of the paths, while a positive rate an increase in path length. To limit the impact of small time-deviations in the measurements, we only consider ASes to which we have traces covering a period of at least two years. To be conservative, we choose to depict change rate per year, but to improve stability of results we require that the samples cover at least twice this span, \ie two years.

Figures~\ref{fig:linreg-slope-ip-hops-mean-cdf} and \ref{fig:linreg-slope-as-hops-mean-cdf} present CDFs of those rates of change for IP and AS hops, respectively.
Generally, paths that go through IXPs are slightly more skewed to the left (shortening) than the paths without \ixps. 
The slight skew to the left of the CDFs is consistent with the previous observations, that \ixps cause moderate path shortenings (see Figure~\ref{fig:avg-as-hops}). 

We see that the majority of ASes experience a change rate of less than 2.5 IP hops and 1 AS hop per year.
In both cases, roughly 60\% of ASes experience a path shortening whereas the remaining 40\% experience a path length increase.
It seems that this lengthening is, in part, driven by the aggressive growth of new networks in developing regions~\cite{fanou2018exploring}. Interestingly, this trend serves as a counterweight to path-shortening experienced by more local networks. 

\begin{figure}
	\centering
	\begin{subfigure}{0.495\columnwidth}
		\includegraphics[width=\columnwidth]{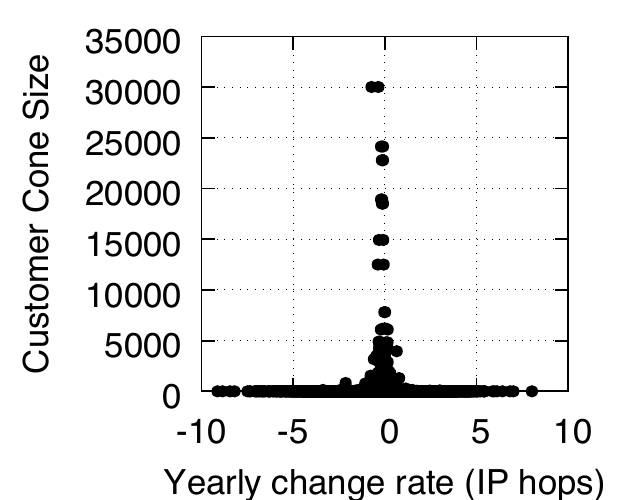}
	\end{subfigure}
	\begin{subfigure}{0.495\columnwidth}
		\includegraphics[width=\columnwidth]{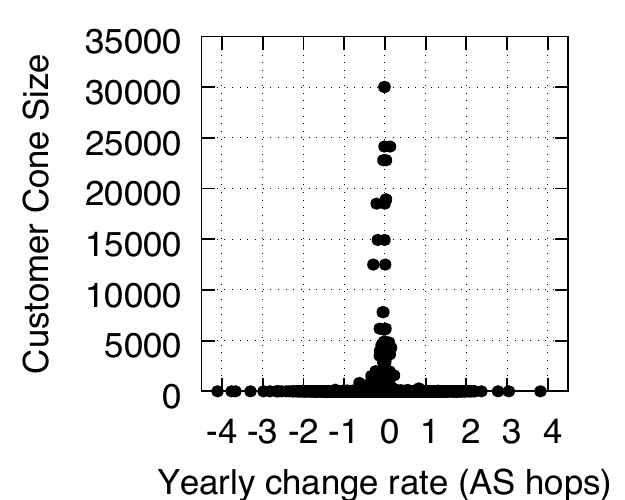}
	\end{subfigure}
	\caption{Change rate per year vs. size of the customer cone of the AS.}
	\label{fig:ascc-vs-slope-ip-vs-slope-as}
\end{figure}

From the CDFs, we also see that ASes experience different rates of path length change over the years.
To get a better understanding, we look at whether the size of the \ases is correlated with the propensity for path length changes.
In Figure~\ref{fig:ascc-vs-slope-ip-vs-slope-as} we plot the change rate an AS incurs against its customer cone size (in terms of \ases).
We see that only smaller ASes (\ie with a small customer cone size) experience a significant deviation in path length.
The larger the size of the customer cone, the smaller the change in path length the AS incurs. We conjecture that since networks that have a large customer cone tend to be close to the core of the Internet, their average path length can hardly be affected. It is the opposite for networks with a small cone size, that tend to be at the edge of the Internet, for which path length can be more easily changed.


\subsection{Path Composition?}

After quantifying how \ixps have enabled path shortening (typically moderate, yet strong in the case of routes towards hypergiants), we next turn our attention to the composition of such routes, \ie the ASes that make up these paths. With a continued increase in the number of peering links, previous researchers have referred to a  \textit{``flattening''} of the Internet structure. Via our historical traceroute data, we strive to bring nuance to this assertion.

\begin{figure}
	\includegraphics[width=\columnwidth]{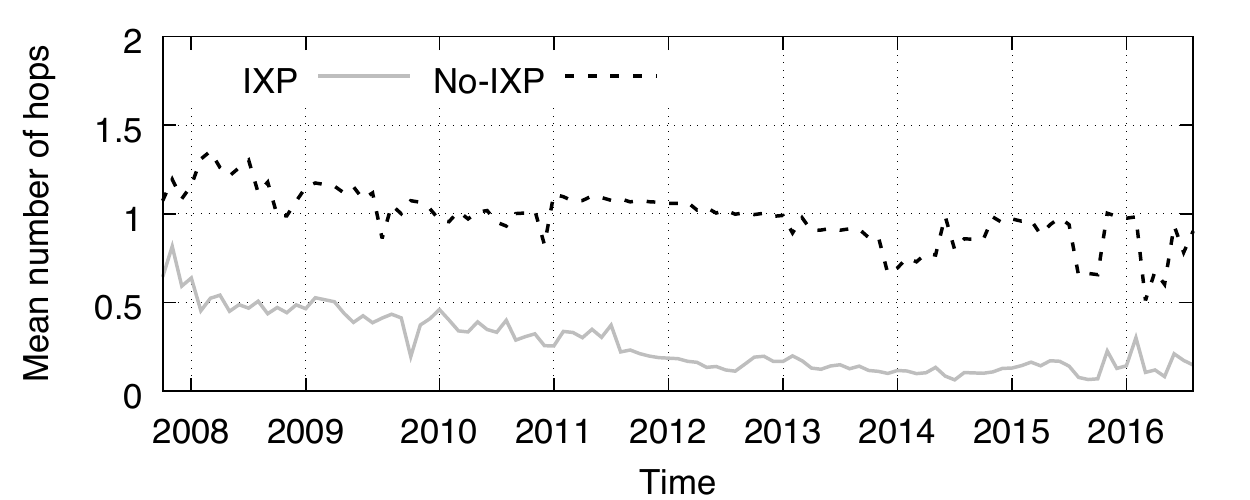}
	\caption{Average number of T1s on the traces.}
	\label{fig:mean-number-t1-ases}
\end{figure}

\paragraph{Path Dependency on Tier-1s.}
First, we assess the dependency of our traceroutes on Tier-1s ove rtime.
Figure~\ref{fig:mean-number-t1-ases} shows the average number of Tier-1s per traceroute and per snapshot, split by whether traceroutes cross an \ixp. We make two main observations:
\textit{(1)} The number of Tier-1s on the paths decreases over time (regardless of the presence of \ixps), which points to a change in the routing hierarchy whereby Tier-1s become less prevalent and can seemingly be replaced by other networks or direct interconnections~\cite{gill2008flattening}.
\textit{(2)} The number of Tier-1s traversed by traceroutes going through IXPs is significantly lower than by those not traversing an IXP.
Indeed, at the beginning of the analysed period, we observe a Tier-1 on almost two or three (iPlane vs. Ark) out of four traceroutes, while by the end of our measurement data we observe that only one in ten traceroutes pass through a Tier-1.
This is aligned with the notion of flattening: given the very low number of Tier-1 ASes on through-IXPs path identified in the most recent snapshots, \ixps have succeeded in allowing a great number of \ases to reduce their dependence on T1 providers.

\begin{figure}
	\includegraphics[width=\columnwidth]{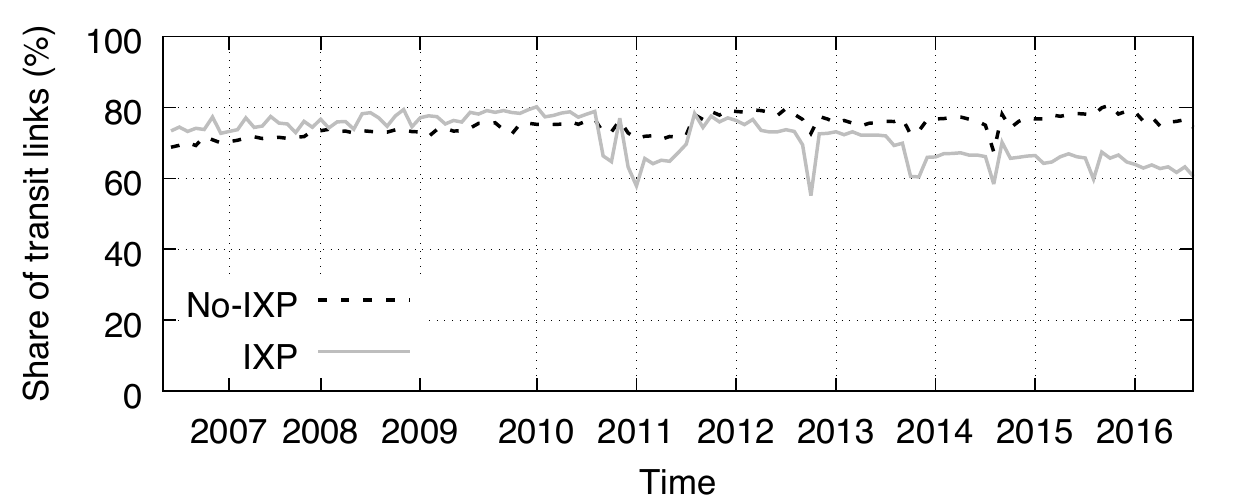}
	\caption{Evolution of the share of transit links.}
	\label{fig:link-types-boxplot}
\end{figure}

\paragraph{Path Relationship Types}
After observing how \ixps have helped in bypassing Tier-1s, we now look more generally at the types of inter-AS relationship that dominate the paths, as well as how \ixps affect path composition. As discussed in Section~\ref{sec:data}, we leverage CAIDA's classification of AS links into peering and customer-provider links.
We then count the occurrences of each type of link in the traceroutes and calculate the relative frequency for traces traversing IXPs and traces traversing no IXP separately.
Figure~\ref{fig:link-types-boxplot} depicts the evolution of the share of transit links (customer-provider type). In the case of traceroutes not traversing an IXP, we see a relatively small share of peering links. Instead, the majority are transit links, in particular customer-provider followed by provider-customer links. For the traces traversing an IXP on the other hand, transit links are still relevant, but in comparison a larger share of links are of the peering type. This reinforces the claim that IXPs have the ability to divert traffic away from paid-for transit links to peering links.
Interestingly, the share of transit links reduces around the same years where hypergiants such as Google or Netflix roll started to expand globally their infrastructure~\cite{bottger2018open,calder2013mapping}

\paragraph{Path Heterogeneity}
The previous analysis has shown that paths consist of a variety of AS types and relationships, and that trends certainly differ for paths traversing \ixps. 
To further explore this, we look at the heterogeneity of \ases as measured by two additional metrics: \emph{(1)} Centrality: the relatively importance of each \as in transferring traffic; and \emph{(2)} Size: the customer cone of each \as along a path.

We start by exploring the centrality of networks, which is defined as the fraction of traces traversing that AS~\cite{sizemore2018dynamic}. In the early (hierarchical) Internet it seems natural to assume that a significant share of traces would pass through Tier-1s, resulting in high centrality values for those networks. With the increasing availability of peering links that cut through the hierarchy, it seems equally expected that traces should be spread more evenly across a larger amount of interconnections. 
We next explore the size and centrality of \ases within our traceroutes, focusing on the impact of \ixps.

\begin{figure}
	\includegraphics[width=\columnwidth]{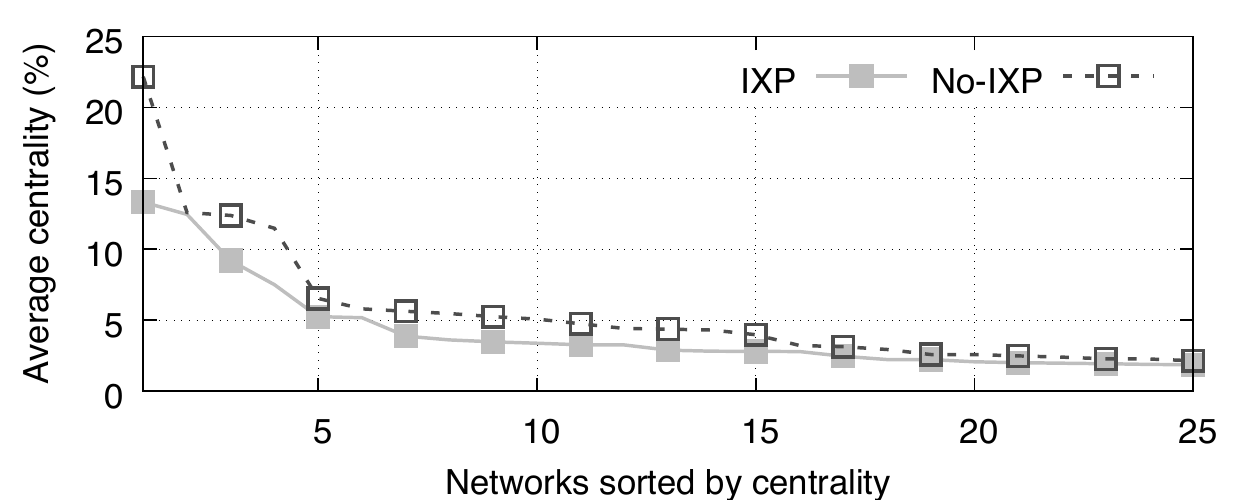}
	\caption{Average centrality over all snapshots for the most central networks.}
	\label{fig:centrality-mean-all-snapshots}
\end{figure}

Figure~\ref{fig:centrality-mean-all-snapshots} presents the centrality averaged over all snapshots for the 25 most central networks (for IXP and non-IXP traces).
As expected, for the traces that do not traverse an IXP, we observe a few rather central networks and then a long tail of \ases with low centrality. 
Interestingly, we  observe a similar picture for the traces going through an IXP, with only slightly lower centrality values than for the traces not traversing an IXP.
This discards trends towards a peering mesh full of direct connections from source to destination networks~\cite{dhamdhere2010internet}.
While we have shown how \ixps are diverting traffic away from Tier-1s (see Figure~\ref{fig:mean-number-t1-ases}), we find that \ixps have not eliminated hierarchical relationships: even at \ixps there is a subset of highly central networks serving as intermediaries for other \ases.

\begin{figure}
	\includegraphics[width=1.05\columnwidth]{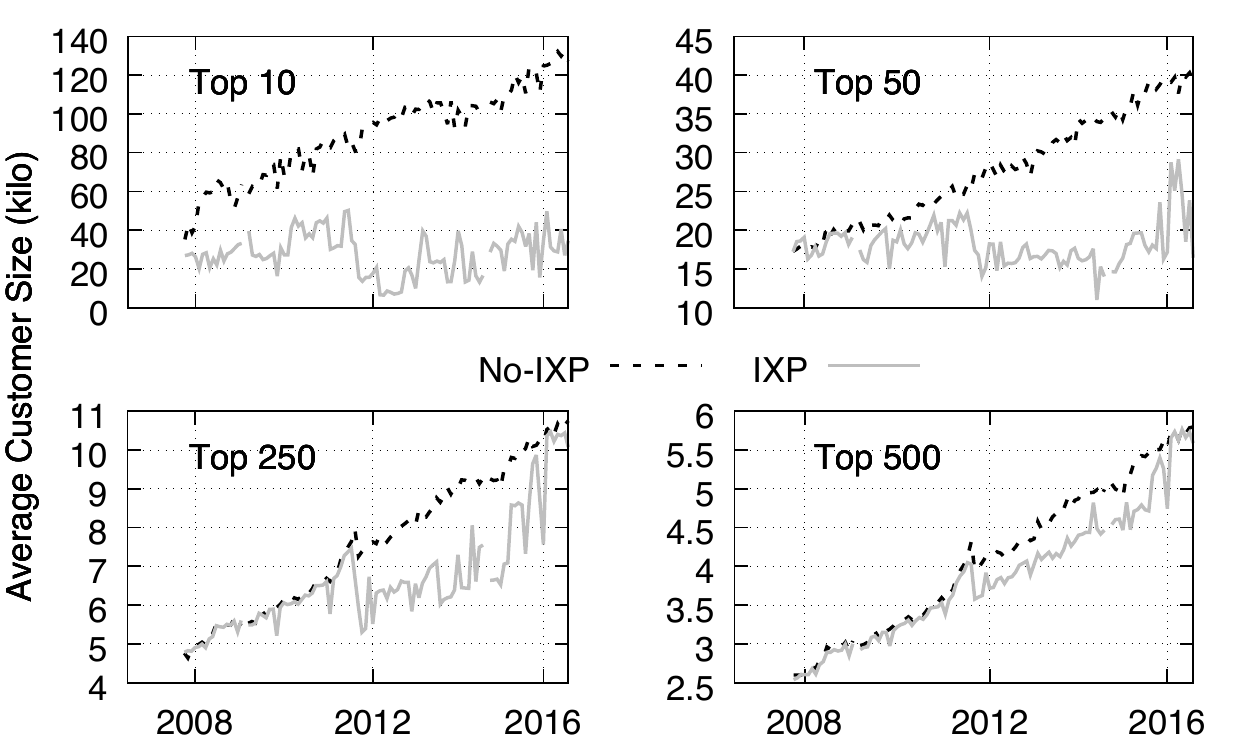}
	\caption{Evolution of average customer cone size for different sets of the top most central networks.}
	\label{fig:asccsize-grid}
\end{figure}

The above observations implies a change in the players building the Internet hierarchy. To explore this, we inspect the 
size of the customer cone of each network as an indication of its role within the Internet hierarchy.
Figure~\ref{fig:asccsize-grid} presents the average customer cone size in terms of the number of ASes, for different subsets of networks according to their centrality.
We observe a trend towards an increasing divergence in customer cone sizes over time, which grows with the centrality of the \as.
For the Top 10 networks, their average customer cone size is bigger for non-IXP traces in comparison to IXP traces, and the difference increase over time.
For the top 25 and Top 250 networks, these differences, while less pronounced, are still present.
The divergence, however, tends to vanish for the top 500 networks.

\begin{figure}
	\includegraphics[width=\columnwidth]{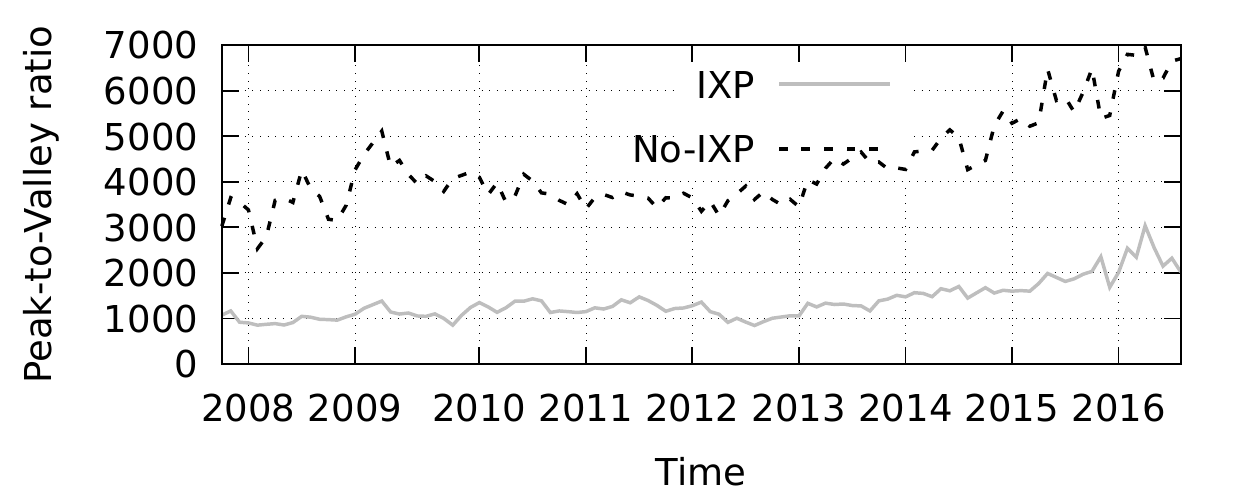}
	\caption{Evolution of the average peak-to-valley ratio of the \ases in each trace, calculated as the ratio of the largest-to-smallest customer cone size in each trace.}
	\label{fig:peak-to-valley}
\end{figure}

This above shows that larger and central \ases (which used to rely on \ixps) have tended to avoid \ixp facilities which have been in turn embraced by smaller \ases. We posit that this may have an impact on the composition of paths: those traversing \ixps may consist of smaller \ases, which are more homogeneous. To check this, we calculate the peak-to-valley ratio for each neighbouring AS within our traceroute data. 
More formally, we compute the ratio of the largest to the smallest \as (in terms of the number of \ases in the customer cone) for each traceroute.
Figure~\ref{fig:peak-to-valley} shows the evolution of the peak-to-valley ratio for traces with and without \ixp.
As expected, it shows that traces traversing \ixps show a lower ratio than those that don't, \ie paths that traverse \ixps tend to contain more homogeneously sized ASes.
Interestingly, however, the ratios diverge over time, showing that \as paths interconnected outside of \ixps tend to have more heterogeneous sizes, particularly in recent years. Of course, this is natural as these paths tend to include very large Tier-1 operators. 
Combining the above results, we see that the composition of paths (measured by size and centrality of ASes) does differ between \ixp-enabled and non-enabled paths. 
That said, the results also shows that, despite of the path-shortening effect and the bypassing of Tier-1s that \ixps brought to the Internet, we still observe a system in which a small number of networks plays a central role.




\subsection{Summary}

This section has shown that \ixp growth has had a clear and evolving impact on the Internet structure and the routes traversing it.
\ixps have enabled a moderate reduction in path lengths for most networks, alongside large reductions for the routes to hypergiants (responsible for a large fraction of the traffic in the Internet).
We find that \ixps have led to a clear bypassing of Tier-1s and more generally reduced the dependency on transit providers.
At the same time, the usage of \ixps has evolved over time where small and less central \ases have replaced large and central \ases in the use of these facilities.
Despite this, we still observe a small set of highly central networks, which have exploited their presence at \ixps to dominate paths, even though they do not adhere to the traditional definition of Tier-1s.

%% file: sections/related.tex
\section{Related Work}\label{sec:related}
The academic community has slowly acknowledged the relevance of \ixps.
Traceroute measurements revealed a sheer number of links traversing this facilities, showing its fundamental role in the Internet~\cite{augustin2009ixps}.
Dissecting one of the largest \ixps~\cite{ager2012anatomy} 
exposed to the research community a rich ecosystem where a wide variety 
of networks exchange large traffic volumes.
Further research revealed large economic benefits of \ixp membership~\cite{castro2014using}
as well as its impact om quality and performance~\cite{formoso2018deep,fanou2016pushing}.
Accordingly, \ixps became the ideal location to place content as close as possible to the users, making this facilities a cornerstone  of the deployment strategy of large content providers such as Netflix~\cite{bottger2018open}. 

Those advantages led to a rapid proliferation of \ixps and \ases therein~\cite{chatzis2015quo}, resulting in a large variety of \ixps in terms of size and business and/or organizational models~\cite{chatzis2013there}.
In addition to this proliferation of peering facilities, its geographical reach was amplified thanks to the emergence of remote peering providers, which allowed far away \ases to peer at the \ixp via Layer-2 connectivity~\cite{castro2014remote,nomikos2016traixroute}.  
By bringing together such a large number of stakeholders and traffic, \ixps have become the target of innovations such as Software Defined Networking~\cite{gupta14,chiesa2016inter,antichi2017endeavour}. 

Aligned with the rising relevance of \ixps, large networks expanded their 
geographical coverage~\cite{krishnan2009moving,bottger2018looking} and 
 peering subsequently increased~\cite{dhamdhere2011twelve}.
 By allowing \ases to  circumvent transit providers~\cite{labovitz2010internet}, these factors pointed to a
a flattening of the Internet topology~\cite{gill2008flattening}, where networks dependence on transit providers decreased. 

In this paper we complete the understanding of the evolution of the Internet 
by enriching its analysis with an \ixp perspective.
Indeed, despite of this startling growth, little is known about the evolution of \ixps, 
the increasing dependence of the Internet on these facilities and its overall impact on the Internet structure.
Some works have taken a look at the Internet evolution as a whole~\cite{dhamdhere2011twelve,labovitz2010internet}, 
but without focussing on these infrastructures. Others, such as Cardona et 
al.~\cite{cardona2012history}, analyse the temporal evolution of an \ixp.
However, its findings for  one specific \ixp cannot be easily extrapolated to a complex and changing ecosystem.

While inferring the Internet structure and its evolution is crucial for 
understanding the risks and bottlenecks~\cite{Giotsas2017outages,dhamdhere2010internet,trevisan2018five}, 
gauging at its complexity is challenging and a variety of sources is necessary.
In understanding peering and \ixps dynamics, PeeringDB has been shown to be 
a reliable data source~\cite{lodhi2014using,bottger2018looking}, which we also use in this paper. 
Similarly, and despite its limitations~\cite{mao2003towards}, traceroute 
repositories, such as iPlane~\cite{madhyastha2006iplane} and Ark~\cite{caida-traceroute}, 
by complementing each other~\cite{huffaker2012internet} can provide a 
representative picture of the Internet~\cite{shavitt2009quantifying}, as we show in this paper.
Aware of the shortcomings of a layer-3 perspective of the Internet~\cite{castro2014remote,willinger2013internet} 
and the incompleteness of the observed Internet topology~\cite{shavitt2009quantifying}, 
we are confident that the usage of multiple datasets of traces and the 
thorough sanitisation and validation conducted mitigate this issue. 
Additionally, identifying \ixps in such traces~\cite{nomikos2016traixroute} 
as well as adding \ixps to the \as-level perspective of the Internet, helps 
in completing the picture. 
Hence, to the best of our knowledge this is the most comprehensive overview of \ixp evolution and their impact to-date. 


%% file: sections/conclusions.tex
\section{Conclusion}\label{sec:conclusions}
Peering and \ixps have reshaped the Internet over the last decade. 
In this paper we have provided a comprehensive analysis of the evolution of the \ixp ecosystem and its impact on the Internet.
Using a rich historical dataset on \ixps and its membership, as well as, two different historical data sets of traces, we identify how the specific impact of \ixps in shaping the Internet has evolved.
We find that while the number of \ixps and its member has more than tripled in the period \period, the share of the address space reachable through them has stagnated.
We also show that \ixps have resulted in substantial path-length shortening in the routes to very large networks, albeit the impact of \ixps on the path-lengths for the rest of destinations has lead to limited path reductions.
Furthermore, we find that whereas large and central \ixps have steadily moved away from public peerings, smaller and less central ones have increasingly relied in this public facilities.
This provides critical insight into how \ixps can be used to improve interconnectivity --- we argue that this is particularly important for regions (like Africa), which are currently undergoing a strategic expansion of \ixps~\cite{fanou2018exploring}. 
Although our study provides comprehensive insights, we also acknowledge certain limitations. 
Most notably, we have little insight into the volumes of traffic traversing the paths we observe. Our future work will involve attempting to understand traffic volumes, as well as other key path characteristics (\eg delay) that may be impacted by \ixps. Naturally, we also intend to broaden our geographical and temporal coverage by incorporating further datasets. 
We will make our datasets and code open for the community to utilize and contribute.
